\documentclass{vgtc}                          % final (conference style)
\ifpdf%                                % if we use pdflatex
  \pdfoutput=1\relax                   % create PDFs from pdfLaTeX
  \usepackage{graphicx}                % allow us to embed graphics files
  \DeclareGraphicsExtensions{.pdf,.png,.jpg,.jpeg} % for pdflatex we expect .pdf, .png, or .jpg files
\else%                                 % else we use pure latex
  \usepackage{graphicx}                % allow us to embed graphics files
  \DeclareGraphicsExtensions{.eps}     % for pure latex we expect eps files
\fi%

\graphicspath{{figures/}{pictures/}{images/}{./}} % where to search for the images

\usepackage{microtype}                 % use micro-typography (slightly more compact, better to read)
\PassOptionsToPackage{warn}{textcomp}  % to address font issues with \textrightarrow
\usepackage{textcomp}                  % use better special symbols
\usepackage{mathptmx}                  % use matching math font
\usepackage{times}                     % we use Times as the main font
\usepackage{cite}                      % needed to automatically sort the references
\usepackage{tabu}                      % only used for the table example
\usepackage{booktabs}                  % only used for the table example
\usepackage{caption}
\usepackage{color}
\usepackage{paralist}
\usepackage{enumitem}
\onlineid{0}

\definecolor{mscolor}{rgb}{0,0,0.8}

\newcommand\toolname[0]{MolecuSense}
\vgtccategory{Research}

\vgtcinsertpkg

\usepackage{url}
\title{\toolname: Using Force-Feedback Gloves for Creating and Interacting with Ball-and-Stick Molecules in VR}

\author{Patrick Gebhardt\thanks{pattigebhardt@gmx.de}\\
    \scriptsize University of Stuttgart
\and Xingyao Yu\thanks{Xingyao.Yu@visus.uni-stuttgart.de}\\%
    \scriptsize University of Stuttgart
\and Andreas Köhn\thanks{koehn@theochem.uni-stuttgart.de}\\%
    \scriptsize University of Stuttgart %
\and Michael Sedlmair\thanks{Michael.Sedlmair@visus.uni-stuttgart.de}\\
    \scriptsize University of Stuttgart}

\abstract{We contribute MolecuSense, a virtual version of a physical molecule construction kit, based on visualization in Virtual Reality (VR) and interaction with force-feedback gloves. Targeting at chemistry education, our goal is to make virtual molecule structures more tangible. Results of an initial user study indicate that the VR molecular construction kit was positively received.  Compared to a physical construction kit, the VR molecular construction kit is on the same level in terms of natural interaction. Besides, it fosters the typical digital advantages though, such as saving, exporting, and sharing of molecules. Feedback from the study participants has also revealed potential future avenues for tangible molecule visualizations.} % end of abstract

\CCScatlist{
  \CCScatTwelve{Human-centered computing}{Human computer interaction (HCI)}{Interaction paradigms}{Virtual reality};
  \CCScatTwelve{Applied computing}{Physical sciences and engineering}{Chemistry}{}
}

\begin{document}


\firstsection{Introduction}

\maketitle

%% \section{Introduction} %for journal use above \firstsection{..} instead
%For education in chemistry it is of broad interest to manually model molecules in order to understand complex spatial structures \cite{andrae_medizin_2013}. Traditionally, these models are created as physical ball-and-stick molecules. Due to physical limitations, these spatial structures can usually only be represented in two dimensions \ms{really?}, which leads to a loss of information that limits the user's freedom of action. In addition, these models are hard to share, replicate, and become fragile at scale.

%For education in chemistry, it is of broad interest to manually model molecules in order to understand complex spatial structures \cite{andrae_medizin_2013}, which usually can be done both physically and digitally. 
3D models of molecules play a fundamental role in chemistry, both in chemical education and in the frontier research, as the spacial structures significantly aid the understanding of molecular reactivity and properties. Traditionally, in the pre-digital ages, physical models, often implemented as ball-and-stick models, have been used extensively. In the recent two decades, this way of representing molecular models was more and more superseded by computer graphics approaches. 

Nevertheless, the traditional physical model bears a lot of advantages, being literally more tangible and offering direct structural manipulations like rotations around bonds which in a natural way allow to explore the conformational space of the molecules. On the other hand, the physical models have some limitations, they become fragile and hard to handle at some larger scale, are difficult to store, to replicate or to share over large distances. %Clearly, they also do not provide any direct link to computational models. 
%Physical models provide haptic feedback and physical perception, helping users understand the structures and increase engagement \cite{howes2018sensing}. However, physical models are naturally limited by the lack of components, and are not conducive to transportation and preservation. 
On the contrary, the molecular construction software can avoid these disadvantages, and allow to precisely manipulate the position and conformation of the atom cluster to create the structures that are more in line with expectations.
%So in the recent two decades, this way of representing molecular models was more and more superseded by computer graphics approaches, which can avoid these disadvantages. 
However, mouse-based interaction and the 2D screens might hinder spatial and haptic thinking in 3D molecular construction tasks.

Ideally, we would like to get a tool, that combines the benefits of these two approaches, by combining tangible experience creating these models with the digital amenities.
%does not sacrifice too much of the tangible experience creating these models, while allowing to leverage the digital amenities.
In collaboration with chemistry researchers, we thus decided to set out to explore how Virtual Reality (VR) and force-feedback gloves could be used as an alternative for creating and interacting with ball-and-stick molecules. 
%The present work deals with the question, whether the newly developed program allows a real three-dimensional experience and a more natural manipulation of the molecular models compared to common PC programs. \todo{what is research gap on your thesis paper? 3D visualization and haptic feedback?}

To test this idea, we built a VR-based prototype using 
the HTC VIVE for output, and force-feedback SenseGloves as input. The SenseGloves \cite{SenseGlove} enable users to grasp virtual objects by controlling motors on the finger joints and thus have the potential to offer a better presence than ordinary VR controllers~\cite{kreimeier2019evaluation}.  While there exists a plethora of molecule visualization in VR \cite{martino2020chemical,cassidy2020proteinvr}, we hypothesized that offering such force-feedback could make for an increased usability and user experience, as existing approaches using stylus-based feedback systems did~\cite{bolopion2009haptic,hou2011six,maciel2004multi,stocks2009interacting}. 
The idea is that interactions such as grasping, holding, and manipulating are becoming more natural and comparable to the current physical approach.

In summary, our work makes the following contributions:
\begin{itemize}[noitemsep,topsep=0pt,parsep=0pt,partopsep=0pt]
    \item The design and implementation of the MolecuSense prototype, that leverages VR and force-feedback.
    \item A preliminary usability evaluation of MolecuSense, revealing some insights into the usage of modern force-feedback gloves for virtual ball-and-stick molecules.
\end{itemize}

\section{Background and Related Work}

In addition to the classic molecular construction kits, in which users assemble the parts by hand, there are now also some computer programs for modelling molecules.
With these programs, not only simple molecules can be formed, but far more complex calculations and visualizations can be done depending on the focus. Avogadro \cite{avogadro_2012}, for example, an open source program, supports the dynamic loading of plug-ins and can therefore always be expanded, so that new types of visualization can be viewed.
Whereas traditional molecular visualizers show molecules in front of you, Molecular Rift, a VR molecule construction kit presented by Norrby et al. \cite{MolecularRift}, lets the user enter the the interior of molecules and explore the unique and realistic 3D effects offered by the VR. They developed a gesture-controlled VR program to easily create molecules for drugs development, which demonstrates its potential when investigating various molecular systems, and received very positive feedback from several focus groups.

One possibility to increase immersion in VR systems is the integration of haptic feedback. 
According to Ramsamy et al. \cite{ramsamy2006using}, presence and efficiency are increased when users can feel and touch virtual objects.
The means used to generate haptic feedback range from devices placed on fingertips to exoskeletal gloves.
Exoskeletal gloves, such as the SenseGlove \cite{SenseGlove} used in our study, cannot give a feeling about the mass of the virtual objects, but it enables the users to perceive the shape of virtual objects.%in this application area, the shape of the objects is in the foreground. %However, this type of haptic input devices can still be improved in terms of comfort and performance.
%Shor et al. \cite{DesigningHaptics} successfully improved these weaknesses by developing the SoftGlove. 
The SoftGlove \cite{shor2018designing}, a further development of the SenseGlove, contains an exoskeleton that lies over the hand and a size-adjustable glove with vibration motors attached to it to improve the force feedback, i.e. the forces acting on the individual fingers and the fit. In addition, more stable fingertip caps have been added to distribute the pressure evenly. This prototype shows that immersion in VR can be improved by haptic feedback, but also needs to be refined in many areas.
%MolecuSense combines the advantages of a VR program and the increased immersion offered by the use of SenseGlove to open up new possibilities in molecule design.

%\ms{Most closely related to our work are approaches that seek to integrate haptic and force feedback into the construction of molecules. There has been a series of approaches that have investigated how stylus-based haptic feedback can be used for such endeavors [add references here]. Our goal is to expand this line of research by looking at modern force-feedback gloves as an input device. }

Most closely related to our work are approaches that seek to incorporate haptic and force feedback into the construction of molecules. Historical endeavors into that direction used mostly hand-grasped, stylus-based, grounded devices~\cite{seifi2019haptipedia}.
Such approaches were found to potentially enhance the spatial visualization \cite{stocks2009interacting}. Besides, such haptic devices might provide force-torque feedback \cite{hou2011six}, and amplify the atomic force in the molecular environment for better perception, and eventually help the users find the best docking position~\cite{bolopion2009haptic,maciel2004multi}. %The force-feedback gloves can also simulate the surfaces and shapes of chemical structures \cite{maciel2004multi}, which 
In MolecuSense, we expand this line of research by looking at modern force-feedback gloves as an input device, which provides a physical perception about the shape of atoms and molecules.

%Our goal is to expand this line of research by looking at modern force-feedback gloves as an input device. 

%\cite{bolopion2009haptic}  either the forces applied by the environment or the internal forces.
%\cite{stocks2009interacting} Using a haptic feedback device in combination with molecular graphics has the potential to enhance three-dimensional visualization.
%\cite{hou2011six}force-torque feedback in force simulation when connecting a 

%\ms{This section needs work. At the moment it gives background on force-feedback devices. This is nor necessarily a bad idea, but it could be shorter and needs to be clearer how it relates to our work. Actual related work, i.e. who is working on the same problem, is largely missing. We are not workinng on a new force-feedback glove, so the above is nice as background but related in term of the problem we want to solve. Imho the closest reltated work would be on: (1) who is suing VR for ball and stick molecules, (2) who has used force-feedback for molecules, (3) what reserach has been done on force-feedback gloves w.r.t.~increasing the usability/user experience of what tasks? ... for each of these categories one paragraph. Last sentence says how we differ from that.    }
\begin{figure*}[tb]
 \centering % avoid the use of \begin{center}...\end{center} and use \centering instead (more compact)
 \includegraphics[width=\textwidth]{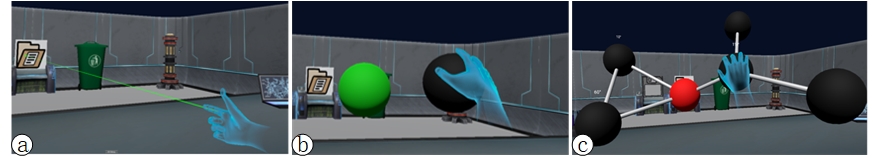}
 \caption{(a) The pointing gesture for operating all basic functions of the program. (b) The users can hold an atom by the grasping gesture. When two atoms are close enough, the unheld one will flash in green, indicating that the current distance can form a bond. (c) When the Edit mode is activated, the selected atom will be marked in red and fixed in space. Then the users can manipulate the unselected atoms and change the length and angles of bonds.}
 \label{fig:MolecuSense}
\end{figure*}
\section{Prototypical Design}

%\ms{overall I recommend to follow the following structure in this section: 1) high-level overview, 2) requirements -- explain physical ball and stick model and say what challenges we are facing when moving that to VR. 3) The visual design of \toolname, 4) the interaction possibilites/gestures, 5) implementation details}

For MolecuSense, we seek to combine digital amenities with the tangible experience of molecule construction. To address that, we visualize the atoms and their chemical bonds with a traditional ball-and-stick model \cite{turner1971ball} in VR. The force-feedback gloves are used to interactively generate and manipulate the molecule models.
%We also allow users to create, delete, and manipulate the atoms to construct the molecules by gestures, which is supported by force-feedback gloves in VR.

%\ms{it might not hurt to introduce the classical physical ball and stick model, show a picture of it, and use it to explain the components: atoms of different types (nodes), connections (edges), vacancies (slots to put in connections, degree of a node) ... it is essentially a 3D node-link diagram, with a specified node degree for each node, right?  would be also good to explicitly state this abstraction .. I suggest to add a subsection 'problem and task characterization' ... also the task analysis from 3.2. can go into this subsection the. Maybe even derive a few requirements at the end  }

\subsection{Design Principles and Requirements}
\label{Sec:Design Principles and Requirements}

%We designed MolecuSense based on the classical ball-and-stick model \cite{turner1971ball} that is widely used in chemistry education. %The molecules visualized based on the ball-and-stick model should carry the following components:

A molecule visualized in the ball-and-stick model consists of atoms (balls) and bonds (sticks) connecting them. The chemical element of each atom is indicated by the ball's color and size; an example is shown in Figure \ref{fig:conditions}(b). The angles between sticks on the same atom should be identical to the bond angles. Abstractly, we can  think of such models as a constrained 3D node-link diagram.  

%visualized following the principles below:
%\begin{compactitem}
%\item \textbf{Nodes} - the basic units representing the atoms, typically spheres. The chemical elements of atoms are indicated by sizes and colors.
%\item \textbf{Sticks} - the connectors between nodes, which represent chemical bonds, The angles between sticks should be identical to bond angles. 
%\item \textbf{Vacancies} - the slots on the nodes to build connections, whose counts are based on the valencies of the atoms.
% \item \textbf{P1}. The atoms are typically represented by spheres. 
% \item \textbf{P2}. The chemical bonds are represented by sticks. 
% \item \textbf{P3}. The angles between sticks on a same atom should be identical to bond angles.
% \item \textbf{P4}. The atoms of different chemical elements are in different colors and sizes.
% \end{compactitem}
Existing graphical user interface of digital tools, such as TmoleX \cite{steffen_tmolexgraphical_2010, furche2014turbomole} and Avogadro \cite{avogadro_2012}, have sought to re-use this ball-and-stick metaphor. By analyzing physical and digital tools, we have summarized a list of basic tasks to interact with ball-and-stick models:
\begin{compactitem}
    \item \textbf{T1}. Creating/removing an atom.
    \item \textbf{T2}. Constructing a molecule by linking the atoms.
    \item \textbf{T3}. Moving/rotating a molecule.
    \item \textbf{T4}. Editing an existing molecule via an operation panel. 
    \item \textbf{T5}. Saving/loading a molecule.
\end{compactitem}

%And one more function for the tools in digital form:
%\begin{compactitem}
    
%\end{compactitem}

\subsection{Visual Design}
\label{Sec:Visual Design}
%MolecuSense employs the ball-and-stick model, where the atoms are represented by 3D balls (typically with a type-specific radius and color) and the chemical bonds by sticks.

In MolecuSense, the atoms are represented by 3D spheres and chemical bonds by sticks, just as in the real physical model. For each atom, we reserved a certain number of vacancies to build bonds, according to the chemical valencies of that atom, which is its capability of forming a certain number of chemical bonds, for instance, four for Carbon and one for Hydrogen. For the atom with multiple vacancies, we preset the position of each vacancy on its surface, to ensure that the single bonds (sticks) connected to this atom form a stable symmetric structure, such as the tetrahedral molecular geometry of Methane (CH$_4$).

In this paper, we focus on Carbon atoms with single chemical bonds for illustrative purposes.

\subsection{Construction of molecules}
\label{sec:Construction of molecules}

To support the tasks outlined above, we designed two gestures and one additional mode for users to construct molecules. 

\paragraph{Pointing gesture}
The virtual pointer has been widely used in distant selection in virtual reality, for instance, gaze \cite{zeleznik2005look} and ray emitted from hand-held controllers. In MolecuSense, we follow this technique and realize it by pointing gesture. When a user makes a "finger gun" gesture, the index finger will emit a ray (Figure \ref{fig:MolecuSense}(a)). To select a target, the user needs to press the thumb towards the index finger. This gesture can be used to click on distant buttons, whose functions include creating/deleting an atom (\textbf{T1}) or opening the folder to save/load (\textbf{T5}) a molecule structure. In addition, when saving a molecular structure, we encoded the position and connection data of the atoms into an XML file \cite{bray2000extensible}.

%\begin{figure}[tb]
% \centering % avoid the use of \begin{center}...\end{center} and use \centering instead (more compact)
% \includegraphics[width=\columnwidth]{Figures/gestures.jpg}
% \caption{(a) The pointing gesture, for operating all basic functions of the program. (b) The users can hold an atom by grasping gesture. When two atoms are close enough, the unheld one will flash in green, indicating that the current distance can form a bond.}
% \label{fig:gesture}
%\end{figure}

\paragraph{Grasping gesture}
The most natural way to interact with 3D objects in virtual environments is to use humans hands \cite{zimmerman1986hand}, which can be supported by force-feedback gloves. In MolecuSense, users can manipulate a molecule by grasping any atom in the molecule, including movement and rotation (\textbf{T3}). The atoms and molecules with electron vacancies can be joined together via drag-and-drop. When a user holds an atom close to an existing atom or molecule and reaches a distance threshold, the existing one will flash, as seen in Figure \ref{fig:MolecuSense}(b). In this case, once the user releases the gesture, the chemical bond represented by a default-length stick will link these two structures (\textbf{T2}).

\paragraph{Edit mode}
When using the pointing gesture to select an atom in an existing molecule, the edit mode will be activated/deactivated. In this mode, the selected atom is marked red and frozen in space, and the angles of all bonds connected to this atom will be presented (Figure \ref{fig:MolecuSense}(c)). The user can manipulate the other atoms via the grasping gesture, to changes the length and angles of bonds, which can even be used to construct ring structures (\textbf{T4}). After releasing the grasping gesture, all unselected atoms will be slightly adjusted in position based on a simple molecular force field with interatomic spring forces.%the interatomic force.

%\begin{figure}
%    \centering
%    \includegraphics[width=\columnwidth]{Figures/apparatus.jpg}
%    \caption{(a)When Edit mode is activated, the selected atom will be in red and fixed in space. Then the users can manipulate the unselected atoms and change the length and angles of bonds. (b) The apparatus in our prototypical system.}
%    \label{fig:Apparatus}
%\end{figure}

%\subsection{Digital Design \ms{Digital Features?}}

%To take full advantage of digital applications, in this system users can save a current structure or load an existing one. After clicking on the "Save" button by Pointing gesture, there will be a virtual keyboard for the users to input the name of the current structure with gloves \ref{fig:keyboard}. The molecular structure data will be encoded into a string and then saved in an XML file. 

%\begin{figure}[tb]
%	\centering % avoid the use of \begin{center}...\end{center} and use \centering instead (more compact)
%	\includegraphics[width=\columnwidth]{Figures/VirtualKeyboard.png}
%	\caption{The virtual keyboard to name and save a created molecule.}
%	\label{fig:keyboard}
%\end{figure}

%After clicking on the "Load" button, the users can select an existent molecule on the list, and then MolecuSense will reconstruct the molecule based on the positional and connection data decoded from the XML file.

%\ms{would be good to restructure this section more into a problem -- solution way, see my comment above.}

\subsection{Implementation}

\begin{figure*}[tb]
 \centering % avoid the use of \begin{center}...\end{center} and use \centering instead (more compact)
 \includegraphics[width=\textwidth]{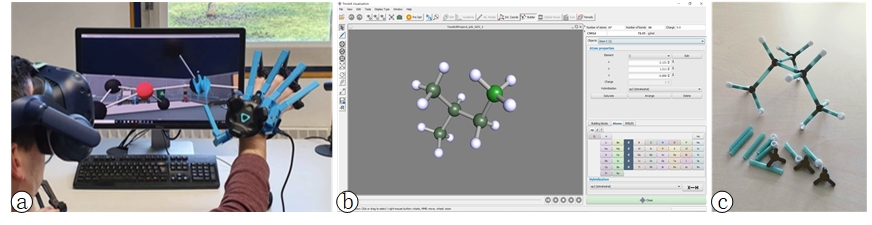}
 \caption{The tools for different conditions in the user study, all based on the ball-and-stick molecular visualization. (a) MolecuSense for VR, where a pair of SenseGloves provides  force feedback. (b) TmoleX for PC. (c) Orbit model kit for PHY.}
 \label{fig:conditions}
\end{figure*}

To implement our technical design, the gloves we use needed to provide force feedback, tactile feedback and hand tracking at the same time. A pair of SenseGloves, which were inspired by an exoskeleton, served as a haptic input device in our prototypical system. The motors on the fingers of the SenseGloves enable it to simulate the grip of the object and the surface type. %Besides, we use an HTC Lighthouse system to track the hand position. 
%Besides, when saving a molecular structure, we encoded the position and connection data of the atoms into an XML file \cite{bray2000extensible}. 
In this study we implemented MolecuSense in a virtual laboratory developed with Unity 3D, and used a HTC VIVE Pro set, which consists of a HMD, two base stations, and two 6-DOF trackers on the gloves (Figure \ref{fig:conditions} (a)). %\ms{move to section 3? As last Subsection 'implementation' there.}

\section{User study}
\label{Sec:user study}

We designed MolecuSense to combine the advantages of digitization and tactile interaction in VR. To verify this, we conducted a user study to compare MolecuSense with existing molecular construction tools that have only one of the above advantages: a PC-based tool, which provides digital amenities, and a physical ball-and-stick construction kit.
While we did not have available a stylus-based haptic device~\cite{bolopion2009haptic}, studying the differences between different haptic devices would be similarly interesting and  poses fruitful avenues for future work.

\subsection{Study Design \& Hypotheses}

A total of 18 people at the  local university participated in this study,  aged 19-39 years (13 males \& 5 females). 12 of them majored in chemistry, while the others had basic school knowledge of chemistry. We conducted our study as a within-subject design, with \textsc{Tool} (VR, PC, physical model (PHY)) as the sole factor. Based on the tasks proposed in Section \ref{Sec:Design Principles and Requirements}, we chose TmoleX \cite{steffen_tmolexgraphical_2010, furche2014turbomole} for the PC condition and prepared an Orbit molecular model kit \cite{Orbit} for the PHY condition (Figure \ref{fig:conditions}). Both tools are also based on the ball-and-stick model. We let the subjects perform a molecule construction task involving carbon atoms, and measured the workload with a questionnaire and task completion time, as well as the usability of different tools.

We formulated two hypotheses for the experiment:
\begin{compactitem}
\item \textbf{H1}. VR and PHY will cause less workload than PC.
\item \textbf{H2}. There is no significant difference between VR and PHY in terms of the system usability.
\end{compactitem}

%\subsection{Study Design}

%As mentioned above, the independent factor in our study is the tool-condition: VR, which is MolecuSense, PC, and PHY. Based on the functional requirements proposed in Section \ref{Sec:Design Principles and Requirements}, we chose TmoleX \cite{steffen_tmolexgraphical_2010, furche2014turbomole} for PC condition and prepared Orbit molecular model kit \cite{Orbit} for the PHY condition (Figure \ref{fig:conditions}). Both tools are based on the ball-and-stick model.

\subsection{Procedure \& Task}

Participants read the study description, and then completed the form of consent and a demographic questionnaire. The participants needed to use different tools to perform the same task. To avoid order effects, we used a Latin Square to counterbalance the order of the tools. The task consisted of the following sequence of steps:%\ms{the following sequence should be called 'the task', which consists of different sub-tasks }

\begin{enumerate}[noitemsep,topsep=0pt,parsep=0pt,partopsep=0pt]
	\item Create the carbon skeleton of 2-methylbutane.
	\item In edit mode, rotate the individual methyl groups in 2-methylbutane into another conformation.
	\item Delete / disassemble the molecule.
	\item Load the prefabricated carbon skeleton of cyclopentane.
	\item Delete at least one atom of cyclopentane and then use the rest to create the carbon skeleton of norbornane.
	\item Save the molecule (only in VR and PC).
\end{enumerate}

After each session, the participants were asked to complete the NASA Task Load Index (TLX) \cite{hart1988development} and the System Usability Scale (SUS) \cite{bangor2008empirical} based on the experience about the current tool. For our VR-based tool, the participants filled also a presence questionnaire \cite{witmer_measuring_1998}. And for each tool, we  recorded the total time that the participants spent on completing the task. After all sessions, we conducted a semi-open interview on subjective preferences and experiences using the different conditions.

%\subsection{Hypotheses}

%We formulated three hypotheses for the experiment:

%\textbf{H1.} The VR and PHY will cause less workload than PC.

%\textbf{H2.} There is no difference between VR and PHY in terms of the System Usability.

\subsection{Results}

Performance data were analyzed using 
%Two-Way Mixed ANOVA, with expertise as the between-subject factor and tool as the within-subject factor.
One-Way Repeated Measures ANOVA, with \textsc{Tool} as the factor. For statistical analyses, we first ran the Shapiro Wilk's test \cite{shapiro1965analysis} for normality of dependent variables. 
%Then we assessed the equality of variances of the dependent variables between different expertise groups using the Levene’s test \cite{levene1961robust}. 
Then we used Greenhous-Geisser corrections \cite{greenhouse1959methods} to adjust the lack of sphericity and reported effect sizes with generalized Eta squared ($\eta_{G}^{2}$) for ANOVA \cite{bakeman2005recommended}. For post-hoc tests, we tested the hypotheses with 95\% confidence interval (CI) and 
%reported them with absolute values 
visualized them respectively \cite{cumming2013understanding}. %Due to the limitation to space, only significant pairwise differences are reported. 
For statistical p-values greater than or equal to 0.001, we report their exact values; for the ones less than 0.001 we report them as ``p\textless.001".

\subsubsection{Completion Time \& Workload}

\begin{figure*}[tb]
    \centering
    \includegraphics[width=.9\textwidth]{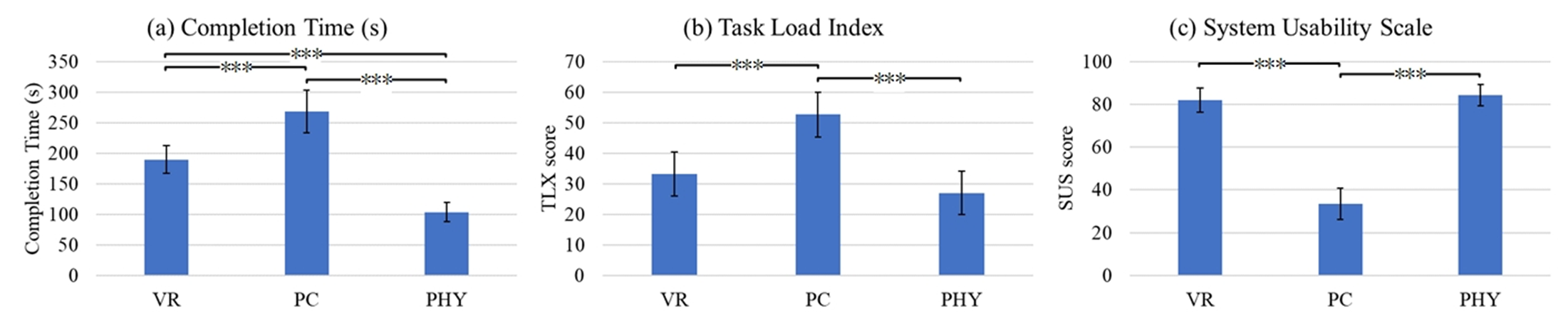}
    \caption{Overview of the results of user study. (a) Completion time: PHY \textless VR \textless PC. (b) TLX: PHY \textless PC \& VR \textless PC. (c) System Usability: PC \textless VR \& PC \textless PHY. The significant differences have been marked with stars * (* for p\textless.05, ** for p\textless.01, and *** for p\textless.001). The error bars denote the 95\% confidence intervals.}
      \label{fig:result}
\end{figure*}

As can be seen in Figure \ref{fig:result} (a), there is a significantly statistical  main effect of \textsc{Tool} on completion time (F$_{2,34}$=67.863, p\textless.001, $\eta_{G}^{2}$=.643).
%=3.08e-12
%presents an overview of completion time depending on expertise and tools. \ms{As can be seen, there is a clear effect between ... bla bla bla} Statistical analysis revealed that there was a significant main effect of Tool (F$_{2,32}$=67.863, p=3.08e-12, $\eta_{G}^{2}$=.655). 
%PHY<VR, (t=6.61, p=2.52e-07, 95\% CI [59.25, 112.20])
%PHY<PC ,(t=9.08, p=3.86e-09, 95\% CI [126.80, 201.53])
%VR<PC,  (t=3.98, p=4.14e-04, 95\% CI [38.19, 118.70])
Overall, PHY had a consistently lower completion time than VR (t=6.61, p\textless.001) and PC (t=9.08, p\textless.001). The completion time for VR was significantly lower than PC (t=3.98, p\textless.001). 

The result of TLX questionnaire showed the same situation (Figure \ref{fig:result} (b)), that the significant main effect was found for \textsc{Tool} (F$_{2,34}$=46.718, p\textless.001, $\eta_{G}^{2}$=.376). PC had a consistently higher TLX score than PHY (t=5.343, p\textless.001) and VR (t=3.99, p\textless.001). The TLX score for VR was trending higher than PHY (t=1.31, p=.20).
%(F$_{2,32}$=42.894, p=8.81e-10, $\eta_{G}^{2}$=.359)
%PC>PHY, (t=5.343, p=6.163e-06, 95\% CI [15.89, 35.40])
%PC>VR, (t=3.99, p=3.35e-04, 95\% CI [9.49, 29.21])

%For the factor Expertise, the normal group had a slight tendency to feel higher workload compared to the group with the high-level chemical background, according to the result of the completion time (F$_{1,16}$=1.551, p=.23, $\eta_{G}^{2}$=.051) and TLX scores (F$_{1,16}$=1.354, p=.26, $\eta_{G}^{2}$=.063).

%\begin{figure*}[tb]
%  \begin{subfigure}[b]{0.33 \textwidth}
%  \centering
%    \includegraphics[width=.9\textwidth]{Figures/Fig.Time.jpg}
%  \end{subfigure}\hfill
%  \begin{subfigure}[b]{0.33\textwidth}
%  \centering
%    \includegraphics[width=.9\textwidth]{Figures/Fig.TLX.jpg}
%  \end{subfigure}
%  \begin{subfigure}[b]{0.33\textwidth}
%  \centering
%    \includegraphics[width=.9\textwidth]{Figures/Fig.SUS.jpg}
%  \end{subfigure}
%  \caption{Overview of the results of user study. (a) Completion time: PHY \textless VR \textless PC. (b) TLX: PHY \textless PC \& VR \textless PC. (c) System Usability: PC \textless VR \& PC \textless PHY. The significant differences have been marked with stars * (* for p\textless.05, ** for p\textless.01, and *** for p\textless.001). The error bars denote the 95\% confidence intervals.}
%  \label{fig:result}
%\end{figure*}

\subsubsection{System Usability}

According to the result of SUS, the system usability was affected significantly by \textsc{Tool} (F$_{2,34}$=110.66, p\textless.001, $\eta_{G}^{2}$=.794). As seen in Figure \ref{fig:result} (c), PC had consistently lower SUS scores than VR (t=11.008, p=\textless.001) and PHY (t=11.94, p\textless.001). And the SUS score for VR was sightly lower than PHY (t=0.66, p=.51).
%However, no significant main effect was found for Expertise to system usability (F$_{1,16}$=.145, p=7.09e-01, $\eta_{G}^{2}$=.0.004).
%(F$_{2,32}$=99.297, p=1.89e-14, $\eta_{G}^{2}$=.784)
%PC<VR  (t=11.008, p=2.51e-12, 95\% CI [39.50, 57.45])
%pc<PHY (t=11.94, p=6.67e-13, 95\% CI [42.14, 59.53])
\subsubsection{Presence}

\begin{figure}[tb]
    \centering
    \includegraphics[width=\columnwidth]{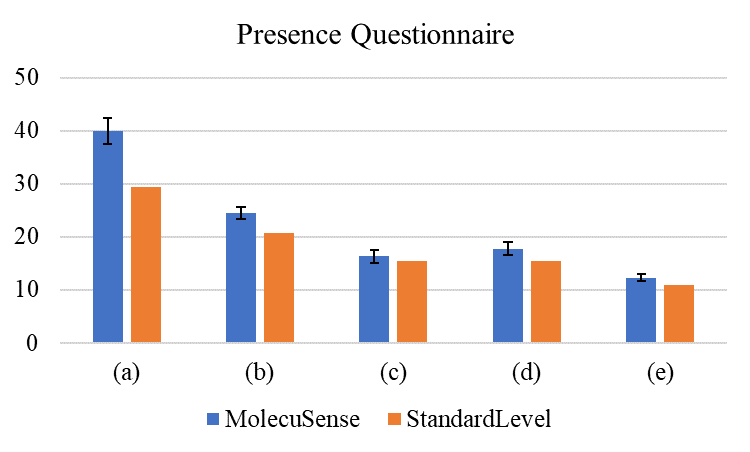}
    \caption{The average values of Presence questionnaire. Error bars show 95\% CIs.  (a) Realism. (b) Possibility to act. (c) Quality of Interface. (d) Possibility to examine. (e) Self-evaluation of performance.}
    \label{fig:presence}
\end{figure}

We assessed the presence of MolecuSence with the average values of the presence questionnaire. As shown in Figure \ref{fig:presence}, we also compared the subscales of the questionnaire with the standard values provided by UQO Cyberpsychological Laboratory \cite{UQO}: (a) Realism, (b) Possibility to act, (c) Quality of Interface, (d) Possibility to examine, and (e) Self-evaluation of performance. Although we do not have the complete dataset for the standard level, it can still roughly be observed, that MolecuSense had higher scores for all subscales than standard level. 
%And the significant differences were confirmed with 95\% confidence intervals except for (c) Quality of Interface.

\subsubsection{Subjective Feedback}

According to the semi-open interviews, 16 out of 18 participants preferred VR compared to the other two tools. The remaining two participants both have professional backgrounds in chemistry. One of them chose PC and the other preferred PHY, because they used the corresponding tools frequently in daily work.

\section{Discussion}

The study results offer preliminary evidence to support our hypotheses H1 and H2.%, and H3.

\textbf{H1 accepted}: Both PHY and VR had lower values than PC in terms of completion time and TLX scale, which supports that these tools caused less workload. According to the subjective feedback, the interactive way of constructing a model by hand felt natural and intuitive, which evoked less mental and temporal demand. 

\textbf{H2 accepted}: We did not find a significant difference between VR and PHY in terms of SUS. Since we take PHY condition as an example of tangible molecule construction, and that VR condition meets the standard levels on most subscales of presence questionnaire, we think this provides evidence that MolecuSense might inherit the advantages of tactile feedback from physical models. Note, that of course with the statistical framework taken, we can never fully prove such a statement.

%\textbf{H3 accepted}: There was no significant main effect of Expertise was found on any scale in our study. We think this also proved the usability of MolecuSense for chemical beginners.

\section{Limitations \& Future Work}

This research has limitations regarding prototype and user study. Many users reported that technical flaws of SenseGlove increased the difficulty of the tasks, although MolecuSense was able to satisfy the basic needs for preliminary construction. For example, the fingertips of the virtual hand did not completely correspond to the real finger, which took the participants more time to perform the gestures. In addition, the inaccuracy of force feedback in some places made the participants miss the best positions, which together with the cable in the glove limited the finger movements.

%In terms of the evaluations, we only measured the task completion time under each condition, and did not measure each task individually, so we did not compare the impact of each tool on different tasks \ms{rephrase as future work ...  at the moment one holistic time measure, future work might break it down into the individual components}.
In terms of evaluation, we only measured the holistic completion time under each condition, so we did not compare the impact of each tool on different sub-tasks. Future work might also break it down into the time metrics for individual tasks. Besides, we did not  include other objective metrics, such as the count of errors when performing sub-tasks. We found that creating a fair error metric is non-trivial, especially under the physical-model condition, in which the participants' errors could only be subjectively perceived by themselves.% \ms{dito, also say why we did not do that in the first place}. 

Although we found initial evidence toward our hypotheses, the molecule construction task in our experiment were relatively simple and only involved Carbon atoms. With more types of atoms and more complex structures involved, the usability and workload of MolecuSense might change.

\section{Conclusion}

%Among existing molecule construction approaches in chemical education, digital desktop tools cannot provide haptic feedback, while physical tools are not conducive to saving and transferring the molecule structures. 
In this paper, we  presented MolecuSense, a VR-based molecule construction tool. This tool combines the advantages of digital amenities and tactile experience, by allowing  users to use force-feedback gloves to manipulate atoms, build bonds, and eventually construct molecules, as well as saving and loading molecule structures. A preliminary study was conducted to compare its workload and usability with existing tools. The results showed that while this VR-based molecular construction kit retains some advantages of digitization, it is closer to the physical models in terms of natural interaction than PC-based tools. %We believe MolecuSense is an efficient, easy-to-learn, and easy-to-use tool and that this technique shows the prospect for tangible approach molecule visualizations.\ms{be more humble}
We hope that our work on MolecuSense will inspire other researcher towards using new forms of interaction for visualization, and by doing so get closer toward the much needed ``science of interaction'' for visualization~\cite{pike2009science}.

\acknowledgments{
This work is funded by Deutsche Forschungsgemeinschaft (DFG, German Research Foundation) under Germany's Excellence Strategy - EXC 2075 - 390740016.}

\bibliographystyle{abbrv-doi-hyperref-narrow}

\bibliography{template}
\end{document}